\newif\ifpeerreview

\ifpeerreview
\documentclass[peerreviewca]{IEEEtran}
\else
\documentclass[conference]{IEEEtran}
\IEEEoverridecommandlockouts
\fi

\usepackage[english]{babel}
\usepackage{amsmath}
\usepackage{graphicx}
\usepackage{textcomp}
\usepackage{xcolor}
\usepackage{tikz} 
\usepackage{pgfplots} 
\usetikzlibrary{pgfplots.groupplots}
\usepackage{float}
\usepackage{longtable}
\usepackage{multirow}
\usepackage{algorithm}
\usepackage{algpseudocode}
\usepackage{booktabs}
\usepackage[marginpar]{todo}
\usepackage{balance}
\usepackage{soul}		
\usepackage{dblfloatfix}
\usepackage{subcaption}
\usepackage{nameref}
\usepackage{notoccite}
\usepackage{hyperref}
\usepackage{paralist}

\def\BibTeX{{\rm B\kern-.05em{\sc i\kern-.025em b}\kern-.08em
		T\kern-.1667em\lower.7ex\hbox{E}\kern-.125emX}}

\graphicspath{{./img/}}

\pgfplotsset{compat = 1.8}

\definecolor{my_green}{RGB}{0,158,115}
\definecolor{my_blue}{RGB}{86,180,233}
\definecolor{my_orange}{RGB}{230,159,0}
\definecolor{my_yellow}{RGB}{240,228,66}
\definecolor{my_red}{RGB}{213,94,0}
\definecolor{my_purple}{RGB}{204,121,167}

\pgfplotscreateplotcyclelist{ColorBlindFriendlyCycleList}{%
	solid, my_green, every mark/.append style={solid, fill=my_green}, mark=x\\%
	solid, my_blue, every mark/.append style={solid, fill=my_blue}, mark=*\\%
	solid, my_orange, every mark/.append style={solid, fill=my_orange}, mark=otimes*\\%
	solid, my_yellow, every mark/.append style={solid, fill=my_yellow}, mark=triangle*\\%
	solid, my_red, every mark/.append style={solid, fill=my_red},mark=diamond*\\%
	solid, my_purple, every mark/.append style={solid, fill=my_purple},mark=*\\%
	dashed, my_green, every mark/.append style={solid, fill=my_green},mark=square*\\%
	dashed, my_blue, every mark/.append style={solid, fill=my_blue},mark=otimes*\\%
}

\pgfplotscreateplotcyclelist{ColorBlindFriendlyCycleListBar}{%
	solid, fill=my_green, every mark/.append style={solid, fill=my_green}\\%
	solid, fill=my_blue, every mark/.append style={solid, fill=my_blue}\\%
	solid, fill=my_orange, every mark/.append style={solid, fill=my_orange}\\%
	solid, fill=my_yellow, every mark/.append style={solid, fill=my_yellow}\\%
	solid, fill=my_red, every mark/.append style={solid, fill=my_red}\\%
	solid, fill=my_purple, every mark/.append style={solid, fill=my_purple}\\%
	solid, fill=black, every mark/.append style={solid, fill=black}\\%
	solid, fill=white, every mark/.append style={solid, fill=white}\\%
}

\algblock{ParFor}{EndParFor}
\algnewcommand\algorithmicparfor{\textbf{parfor}}
\algnewcommand\algorithmicpardo{\textbf{do}}
\algnewcommand\algorithmicendparfor{\textbf{end\ parfor}}
\algrenewtext{ParFor}[1]{\algorithmicparfor\ #1\ \algorithmicpardo}
\algrenewtext{EndParFor}{\algorithmicendparfor}

\usepackage{environ}
\NewEnviron{scaled_IEEEeqnarray}{%
	\begin{IEEEeqnarray}{l}
		\scalebox{0.8}{$\BODY$}
	\end{IEEEeqnarray}
}

\begin{document}
	
	\ifpeerreview
	\title{Efficient Edge AI: Deploying Convolutional Neural Networks on FPGA with the Gemmini Accelerator}
	\else
	\title{Efficient Edge AI: Deploying Convolutional Neural Networks on FPGA with the Gemmini Accelerator \thanks{This research was funded by the German Federal Ministry of Education and Research within the project "GreenEdge-FuE“, funding no. 16ME0517K.}}
	\fi
	
	\author{\IEEEauthorblockN{Federico Nicolás Peccia}
		\IEEEauthorblockA{\textit{FZI Research Center for Information Technology} \\
			Karlsruhe, Germany \\
			peccia@fzi.de}}
	
	\author{
		\IEEEauthorblockN{Federico Nicolás Peccia\IEEEauthorrefmark{1}\IEEEauthorrefmark{2}, Svetlana Pavlitska\IEEEauthorrefmark{1}\IEEEauthorrefmark{3}, Tobias Fleck\IEEEauthorrefmark{1}\IEEEauthorrefmark{3}, Oliver Bringmann\IEEEauthorrefmark{1}\IEEEauthorrefmark{2}}
		\IEEEauthorblockA{FZI Research Center for Information Technology\IEEEauthorrefmark{1}, University of Tübingen\IEEEauthorrefmark{2}, Karlsruhe Institute of Technology\IEEEauthorrefmark{3}\\
			Germany \\
			peccia@fzi.de,pavlitska@fzi.de,tobias.fleck@fzi.de,oliver.bringman@uni-tuebingen.de}
	}
	
	
	
	\ifpeerreview
	\IEEEpeerreviewmaketitle 
	\else
	\maketitle
	\fi
	
	\begin{abstract}
		
		
		
		The growing concerns regarding energy consumption and privacy have prompted the development of AI solutions deployable on the edge, circumventing the substantial CO2 emissions associated with cloud servers and mitigating risks related to sharing sensitive data. But deploying Convolutional Neural Networks (CNNs) on non-off-the-shelf edge devices remains a complex and labor-intensive task. In this paper, we present and end-to-end workflow for deployment of CNNs on Field Programmable Gate Arrays (FPGAs) using the Gemmini accelerator, which we modified for efficient implementation on FPGAs. We describe how we leverage the use of open source software on each optimization step of the deployment process, the customizations we added to them and its impact on the final system's performance. We were able to achieve real-time performance by deploying a YOLOv7 model on a Xilinx ZCU102 FPGA with an energy efficiency of 36.5 GOP/s/W. Our FPGA-based solution demonstrates superior power efficiency compared with other embedded hardware devices, and even outperforms other FPGA reference implementations. Finally, we present how this kind of solution can be integrated into a wider system, by testing our proposed platform in a traffic monitoring scenario.
		
	\end{abstract}
	
	\begin{IEEEkeywords}
		Convolutional Neural Networks, Field-Programmable Gate Arrays, FPGA accelerator, Edge Computing, Deployment Workflow, Energy Efficiency.
	\end{IEEEkeywords}
	
	\section{Introduction}
\label{section:introduction}


Nowadays, several pre-trained Convolutional Neural Networks (CNN) already exist that cover common tasks like object detection or tracking, so the burden of training the models from scratch can be avoided. A lot of commercial edge devices (for example, the Jetson family of NVIDIA) provide software frameworks which take care of optimizing these models for deployment on them.

But taking these pre-trained models and deploying them on an edge device that is \textit{not} an off-the-shelf hardware (for example, a custom hardware accelerator), can be a very manual and time consuming task. This task also requires a wide set of skills to be able to extract as much performance as possible of the proposed hardware-software system, ranging from knowledge about CNN optimization techniques to understanding the internals of the desired hardware architecture. Moreover, the final system cannot exist in isolation, so standard interfaces need to be provided to easily obtain the data to be processed and to publish the output of the CNN to the next system in the processing pipeline, be it another edge hardware or a cloud server.

In this paper, we aim to simplify these deployment activities by introducing an end-to-end workflow based on FPGA for CNN deployment on the edge using the Gemmini accelerator \cite{gemmini-dac}, where each of these aspects is taken into account. We describe each step of the deployment workflow (including hardware-aware model modifications, quantization, layer scheduling optimizations, model partitioning and more) and its impact on the resulting system performance.

We were able to achieve real time performance by deploying the YOLOv7 model \cite{yolov7} on a Gemmini accelerator \cite{gemmini-dac} implemented on a Xilinx ZCU102 FPGA development board. We expanded the Gemmini integration for the TVM Deep Learning compiler presented in \cite{peccia2022integration}, and optimized the execution of the CNN using the AutoTVM framework \cite{autotvm}. We also compare the resulting energy efficiency of the FPGA-based solution against reference implementations (including server-size and embedded GPUs), and against other FPGA accelerators, and extract conclusions from it. We also provide an example use case to show how this kind of solution can be integrated into a wider system.



The rest of this paper is organized as follows. First, Section \ref{section:related_work} presents related works deploying CNN on FPGAs using a variety of frameworks. Then, Section \ref{section:hw} describes the selected hardware accelerator, and the optimizations implemented for it which improve its mapping onto an FPGA. Section \ref{section:workflow} describes the end-to-end software workflow, starting with the pretrained CNN model and explaining each optimization step. Then, Section \ref{section:eval} compares the proposed solution against reference implementations. Section \ref{section:case_study} presents a case study on how the proposed workflow can be used to integrate a CNN into a wider edge system. Finally, Section \ref{section:conclusions} concludes our work.

\section{Related work}
\label{section:related_work}

\begin{figure*}[htbp]
	\centering
	\includegraphics[width=0.9\textwidth]{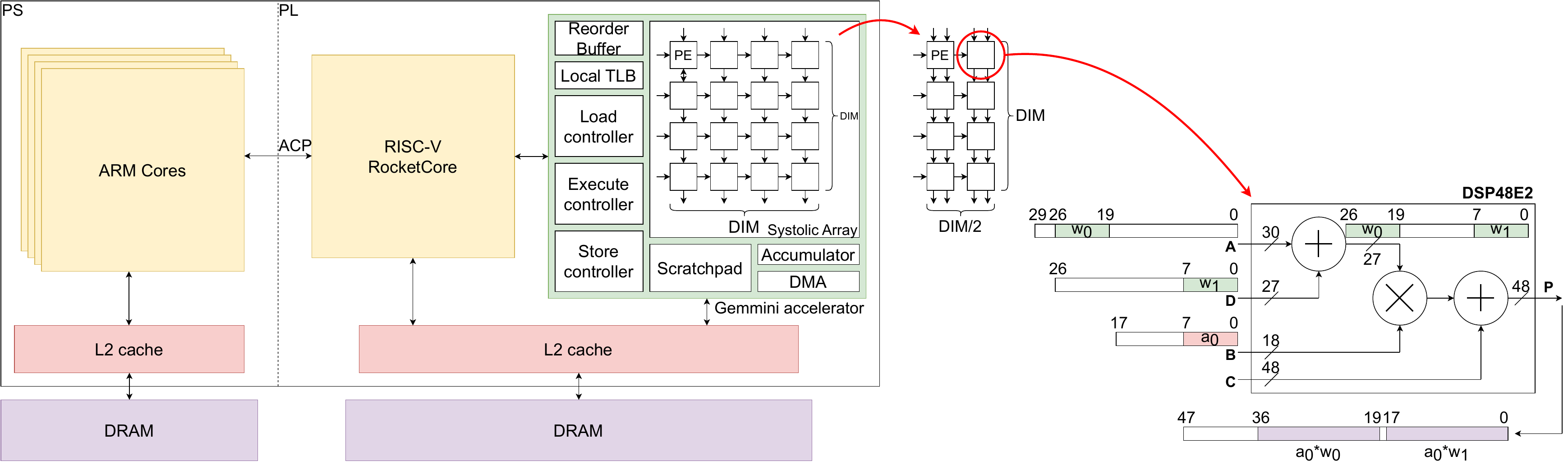}
	\caption{Proposed hardware solution, showing the improvement in the size of Gemmini's systolic array using the DSP packing technique, and a detailed description of how the DSP packing is implemented on the DSP48E2 available on Xilinx FPGAs.}
	\label{fig:soc_architecture}
\end{figure*}

The automation frameworks to map CNNs on FPGAs can be separated in two distinct groups \cite{hamanakaExplorationStateoftheArtAutomation2023,venierisToolflowsMappingConvolutional2018}. Several works have focused on implementing stream or dataflow type accelerators, where each layer is mapped as a separate hardware module on the FPGA and are interconnected in a stream-like manner. These works are able to achieve high throughputs, but the sizes of the CNNs that can be implemented are limited by the available FPGA resources. HLS4ML \cite{aarrestadFastConvolutionalNeural2021} translates a high level description of a Neural Network into High Level Synthesis language (HLS), and uses this new description to implement the needed hardware modules on the FPGA. FINN \cite{blottFINNREndtoEndDeepLearning2018} is a Xilinx tool that generates a complete dataflow type accelerator that is tailor made for a specific Quantized Neural Network (QNN). FlexCNN \cite{basalamaFlexCNNEndtoendFramework2023} also uses HLS, but generates systolic array based accelerators. DnnWeaver \cite{sharmaHighlevelDeepNeural2016} uses hand-optimized hardware templates to create the FPGA accelerator. fpgaConvNet \cite{7544745} provides a tool to generate throughput-oriented HLS for a given model, taking into account the resource constraints of the desired FPGA.

But a stream type accelerator is no longer suitable when a more computation or memory intensive CNN needs to be implemented, because the design can very quickly exceed the available resources of the selected FPGA platform. This is where overlay or single-engine type accelerators present several advantages. These kind of accelerators contain hardware modules which execute typical operations needed by CNNs. The accelerators are independent of the selected CNN, so they can be parametrized to fit the available FPGA first, and then a software framework maps the execution of the different layers of the CNN onto the accelerator. Angel-Eye \cite{guoAngelEyeCompleteDesign2018} presented a CNN accelerator together with a compilation workflow to map a generic CNN onto it. VTA \cite{vta} provides an accelerator integrated into TVM for fast deployment and tuning of CNN onto Xilinx FPGAs. DYNAMAP \cite{liuEndEndFramework2022} takes into account the heterogeneity of layers on a CNN when selecting the right hardware template for the accelerator. Vitis AI \cite{vitis-ai} is a commercially available tool to optimize CNN for deployment on Xilinx Deep Processing Units (DPU).

A software integration to deploy CNNs onto the Gemmini accelerator already exists \cite{githubGitHubUcbbaronnxruntimeriscv}. This uses the ONNX Runtime to parse layers and offload them to the accelerator. But this approach is severely limited. First of all, in order to use it, the ONNX Runtime needs to be compiled for the embedded system, which in certain cases may be a limitation, specially in constrained bare metal scenarios. Second, the current integration is limited only to convolutions and dense layers. Finally, it completely ignores the possibility of exploring the schedule space for each layer, which can greatly improve the performance of the model by changing the order in which the individual instructions are dispatched to the accelerator.

\section{The Gemmini accelerator}
\label{section:hw}

\begin{figure*}[htbp]
	\centering
	\includegraphics[width=\textwidth]{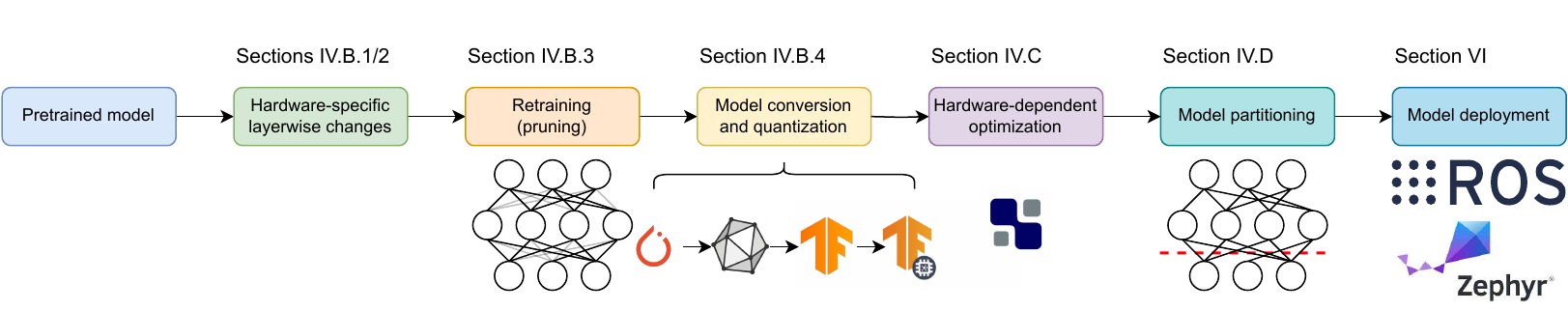}
	\caption{End-to-end deployment workflow}
	\label{fig:workflow}
\end{figure*}

When working with single-engine accelerators, it is important to take into account which features are supported by it, and if these are enough to accelerate the desired CNN. In this work, the selected hardware accelerator is Gemmini \cite{gemmini-dac}, a highly customizable systolic array based accelerator. Because of several factors, including (but not limited to) documentation, flexibility and development status, Gemmini provides an excellent platform for these kinds of projects. It also presents low jitter and latency when compared with other open source FPGA accelerators \cite{10361287}.

The accelerator consists of three decoupled modules: the \textit{Load} controller (which moves chunks of data from the external memory to the internal accelerator scratchpads), \textit{Execute} controller (which dispatches data already available in the scratchpads to the systolic array) and \textit{Store} controller (which moves chunks of data from the accelerator's internal scratchpads to the external memory system). 

Gemmini needs to be attached to a RocketCore CPU \cite{rocketcore}, which orchestrates the execution of the CNN, by offloading layers on the accelerator. The CPU may issue two types of instructions to the accelerator: \begin{inparaenum}[1)]
	\item CISC-type instructions, which use hardcoded state machines inside the accelerator to execute typical operations like tiled matrix multiplications and convolutions, abstracting the developer of the internal operation of the accelerator, and
	\item RISC-type instructions, which give the developer a far more fine-grained control over the working of the accelerator, by providing intrinsic instructions to move data in or out of the accelerator, and execute matrix multiplications using the data available in Gemmini's internal scratchpads or accumulator.
\end{inparaenum}

Although a tape out of the Gemmini accelerator can be realized (see \cite{gemmini-dac}), deploying it onto an FPGA provides an interesting cost-benefit tradeoff. It provides a flexible middle ground where hardware optimizations can be explored, while still providing an energy efficient platform. We selected FPGAs based on Xilinx's Zynq SoC architecture, as these also contain ARM Cores (Processing System, PS) alongside the proper FPGA (Processing Logic, PL). This heterogeneous SoC architecture can be utilized to optimize the model's performance by distributing the execution of the network across both parts of the SoC (see Section \ref{section:model_partitioning}). The FPGA bitstream was realized using a similar workflow to the one provided by the Chipyard project \cite{chipyard}.

\subsection{FPGA related optimizations}
\label{section:FPGA_opt}

The accelerator was optimized for FPGA deployment by implementing the DSP packing technique proposed by \cite{dsp-pack}. DSPs are configurable slices contained in Xilinx FPGAs, which can very efficiently implement multiply-accumulate operations. This is the same mathematical operation used in each core Processing Element (PE) of the Gemmini accelerator. We first modified the accelerator to map each PE onto one DSP. But the DSPs were underutilize, as they can execute wide bit multiplications at its core, and Gemmini's PEs use 8 bits only to describe each input to its multiplier. So we implemented the technique proposed by \cite{dsp-pack}. By doing so, we were able to pack two 8 bit weights multiplications onto the same DSP, and therefore halve the usage of DSP slices.

The FPGA design can be further optimized by disabling a number of modules provided by the Gemmini accelerator, which are not needed for deployment of YOLO-type networks. Normalization features (used for transformer networks), transposition modules, virtual addresses translation tables and kernel dilation capabilities (used in encoder-decoder networks) are all examples of modules that can be disabled in order to reduce the resource usage of the final design. We were also able to reduce the output scaling modules which apply a floating point scaling factor to the accumulated output of the accelerator to reduce the bit width back to 8 bits, by changing the scaling factor from float32 to float16, without appreciating any degradation in the performance of the deployed CNN.	

Figure \ref{fig:soc_architecture} presents the final hardware architecture of the selected SoC, and explains how the DSP packing technique is implemented to modify the structure of the systolic array. Section \ref{section:eval} analyzes the impact of this modifications.

\section{End-to-end worklflow}
\label{section:workflow}

Now that the accelerator is already optimized to fit the selected FPGA platform, Figure \ref{fig:workflow} describes our end-to-end software process to orchestrate the execution of the CNN on the accelerator. As described in Section \ref{section:introduction}, we start with a pretrained CNN model for object detection and the hardware on which we want to deploy it, and we end up with a complete system tuned for the selected hardware.

\subsection{CNN model description}

We selected the state-of-the-art YOLOv7 model \cite{yolov7}, which is a CNN model pretrained on the COCO dataset for object detection. The model is provided in different sizes, so we selected the smaller version, YOLOv7-tiny (6.2 million parameters), as the target model to deploy. 

This model, like other currently state-of-the-art CNN models for object detection, can be separated into two very distinct parts. The first one is the main part of the model, the one containing all the convolutions and other heavy tensor-intensive computation tasks, like concatenation or fully connected layers. The second part is composed of the layers preparing the data for the Non Max Suppression (NMS) algorithm and the algorithm itself, which filters and merges the proposed bounding boxes of the model and outputs cleaner detections. Contrary to other FPGA accelerator works, which only focus on deploying the feature extractor of the CNN (or only the convolutional layers), as we are focusing on a complete end-to-end solution, we evaluate the deployment of both parts of the network, their respective performance, and the optimal mapping to distribute the execution of both parts across the heterogeneous SoC.

\subsection{Model optimizations}

\subsubsection{Input image size selection}

The input image size of a CNN is directly related to the amount of giga computation operations (GFLOP when talking about floating point operations, GOP when referring to integer arithmetic) needed per inference, and thus directly related to the execution latency of the model. But this hurts the quality of the model's output, given that smaller input image sizes also reduce the amount of features that can be detected for each object in the image. So the first step is to select an appropriate input image size.

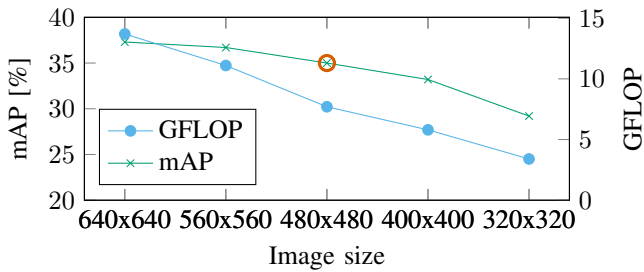
\begin{figure}[t]
	\centering
	\begin{tikzpicture}
		\begin{axis}[
			xlabel=Image size,
			ylabel={mAP [\%]},
			height=0.17\textheight,
			width=0.34\textheight,
			ytick pos=left,
			ymin=20,
			ymax=40,
			xtick=data,
			cycle list name=ColorBlindFriendlyCycleList,
			symbolic x coords={
				640x640,
				560x560,
				480x480,
				400x400,
				320x320
			},
			xtick={
				640x640,
				560x560,
				480x480,
				400x400,
				320x320
			},
			xticklabels={
				640x640,
				560x560,
				480x480,
				400x400,
				320x320
			}
			]
			\addplot+ plot [] table[x=input_image_size,y=mAP,col sep=comma] {./data/input_image_size.csv};
			\label{mAP0}
			\draw[very thick, my_red] (axis cs: 480x480,35) circle[radius=0.1cm];
		\end{axis}
		\begin{axis}[
			ylabel={GFLOP},
			height=0.17\textheight,
			width=0.34\textheight,
			ylabel near ticks,
			ytick pos=right,
			ymin=0,
			xtick=data,
			cycle list name=ColorBlindFriendlyCycleList,
			legend style={at={(0.03,0.50)},anchor=north west},
			legend cell align={left},
			symbolic x coords={
				640x640,
				560x560,
				480x480,
				400x400,
				320x320
			},
			]
			\pgfplotsset{cycle list shift=1}
			\addplot+ plot [] table[x=input_image_size,y=GOP,col sep=comma] {./data/input_image_size.csv};
			\addlegendimage{/pgfplots/refstyle=mAP0}
			\legend{GFLOP, mAP}
		\end{axis}
	\end{tikzpicture}
	\caption{Selection of the input image size}
	\label{fig:image_size_analysis}
\end{figure}

For the YOLOv7-tiny, we analyse the change in the mean Average Precision metric (mAP) for different image sizes for the pretrained model. As can be seen in Figure \ref{fig:image_size_analysis}, the mAP of the model gets worse the smaller the input image size. Given that the mAP remains almost stable until an input image size of 480x480 pixels and then starts to get worse, we select 480x480 as the input image size we are going to use for the rest of the paper, which reduces the number GFLOPS by almost 50 \%.

\begin{table*}[t]
	\caption{mAP [\%] for the evaluated model versions across different frameworks, using $480\times 480$ as image size}
	\begin{center}
		\begin{tabular}{ccccccccc} \toprule
			Model & PyTorch & PyTorch & ONNX & Tensorflow & TFLite-float32 & TFLite-float16 & TFLite-int8 & TVM \\
			& Pretrained & Ours & Ours & Ours & Ours & Ours & Ours & Ours \\
			\midrule
			YOLOv7-tiny & 35.2 & 33.1 & 32.2 & 32.2 & 32.2 & 32.1 & 29.6 & 29.2 \\
			40\% pruned YOLOv7-tiny & - & 30.5 & 30.0 & 30.0 & 30.0 & 30.0 & 27.0 & 26.7 \\
			88\% pruned YOLOv7-tiny & - & 20.8 & 20.6 & 20.6 & 20.6 & 20.6 & 18.4 & 18.4 \\
			\bottomrule
		\end{tabular}
		\label{tab:mAP_per_framework}
	\end{center}
\end{table*}

\subsubsection{Activation function replacement}

The range of layers supported by custom accelerators is usually restricted, and the same applies to the activation layers that can be directly accelerated on the hardware. As such, the Gemmini accelerator does not support the LeakyReLU activation used by the original YOLOv7-tiny model. Instead, these LeakyReLU layers would be mapped to the RISC-V CPU attached to the accelerator, which hurts the execution latency of the deployed model. But the Gemmini accelerator does support the ReLU layer, so we replaced all LeakyReLU layers of the model with simpler ReLU6 activation layers.

\subsubsection{Pruning}

\begin{figure}[t]
	\centering
	\begin{tikzpicture}
		\begin{axis}[
			xlabel={Iteration},
			ylabel={mAP [\%]},
			height=0.17\textheight,
			width=0.32\textheight,
			ytick pos=left,
			ymin=0,
			ymax=35,
			cycle list name=ColorBlindFriendlyCycleList,
			]
			\addplot+ plot table[x=iteration,y=mAP,col sep=comma] {./data/pruning.csv};
			\label{mAP1}
			\draw[very thick, my_red] (0,331) circle[radius=0.1cm];
			\draw[very thick, my_red] (30,305) circle[radius=0.1cm];
			\draw[very thick, my_red] (140,208) circle[radius=0.1cm];
		\end{axis}
		\begin{axis}[
			ylabel={Sparsity [\%]},
			height=0.17\textheight,
			width=0.32\textheight,
			ylabel near ticks,
			ytick pos=right,
			ymin=0,
			ymax=100,
			cycle list name=ColorBlindFriendlyCycleList,
			legend style={at={(0.63,0.50)},anchor=north west},
			legend cell align={left},
			]
			\pgfplotsset{cycle list shift=1}
			\addplot+ plot table[x=iteration,y=sparsity,col sep=comma] {./data/pruning.csv};
			\addlegendimage{/pgfplots/refstyle=mAP1}
			\legend{Sparsity,mAP}
		\end{axis}
	\end{tikzpicture}
	\caption{Selection of the pruned models based on mAP and parameter sparsity}
	\label{fig:pruning}
\end{figure}
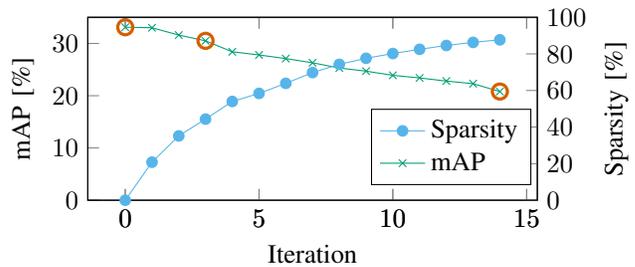

Model parameter size plays a vital role in the deployment of CNNs on memory-constrained devices. Pruning aims at removing redundancy in large models. Unlike unstructured pruning, where certain weights or parameters are replaced with zeros, structured pruning aims at deleting structural elements of a network, like neurons or filters. For CNNs, filter pruning is usually applied. The architecture of YOLOv7 is very complex; a lot of concatenation layers make filter pruning challenging. We pruned the CNN using iterative pruning \cite{pavlitska2024iterative}\footnote{\url{https://github.com/fzi-forschungszentrum-informatik/iterative-yolo-pruning}}, which relies on a connectivity graph of convolutional layers. In each iteration, layers to be pruned as well as a pruning rate are selected. Then, the model is pruned and fine-tuned to compensate for a drop in accuracy. We could thus achieve a reduction of up to 88\% of parameters and up to 78\% of GFLOPS after 14 iterations at a cost of a drop of 12.3 percent points in mAP (see Figure \ref{fig:pruning}).

A trade-off between mAP and parameter sparsity during pruning should be taken into account depending on the application. For the rest of the paper, we choose to evaluate the original un-pruned model and also two pruned models: the pruned model with 40\% sparsity (as an example of a model where the mAP was not degraded below 30\%) and the one with 88\% sparsity (to show the minimum latency we can achieve with an extremely tiny model).

\subsubsection{Framework conversion}
\label{section:frameworks}

\begin{table*}[t]
	\caption{Resource consumption of implemented FPGA accelerators}
	\label{tab:hw_comparison}
	\begin{center}
		\begin{tabular}{ccccccccc} \toprule
			Accelerator & Board & Frequency [MHz] & LUT & FF & BRAM & URAM & DSP & LUTRAM \\
			\midrule
			Gemmini (Original) & ZCU102 & 100 & 133376 & 103026 & 613 & 0 & 441 & 11181 \\
			Gemmini (Ours) & ZCU102 & 150 & 150596 & 122028 & 693 & 0  & 652 & 11225 \\
			Gemmini (Ours) & ZCU111 & 167 & 156413 & 134787 & 321.5 & 78 & 652 & 13064 \\
			VTA (Ours) & ZCU111 & 100 & 37616 & 10924 & 70 & 12 & 0 & 2982 \\
			\bottomrule
		\end{tabular}
	\end{center}
\end{table*}

To map the model to different hardware backends using the TVM framework, we first need to apply several model conversions across multiple frameworks. 

First, we export the PyTorch model to ONNX using its standard ONNX exporter. Then, onnx2tf \footnote{\url{https://github.com/PINTO0309/onnx2tf}} is used to transform the model to Tensorflow, which takes care of changing the data layout format NCHW (used by PyTorch and ONNX) to NHWC, which is the layout supported by the next step in the process.

We used the TFLite framework to quantize the model to \textit{int8} representation \cite{DBLP:journals/corr/abs-1712-05877}. TFLite provides the option to use a different scaling and offset factor for each channel of each tensor (per-channel quantization), which reduces the quantization error, but we chose to use a per-tensor quantization for ease of deployment on the Gemmini accelerator. We also excluded from the quantization the Non Max Suppression (NMS) process at the output of the model, as our empirical tests have demonstrated a significant loss in prediction quality when quantizing this part of the model. Finally, the quantized model is imported into TVM using its standard TFLite frontend.

But transforming a model from one framework to the other can be hurtful for the quality of the predictions, particularly when using quantization procedures. In order to validate each conversion step, Table \ref{tab:mAP_per_framework} presents the mean average precision (mAP) of each model after the conversion to each framework. We also report as a baseline the mAP of the pretrained model provided by the authors of the original YOLOv7 paper. As expected, quantizing the model degrades the performance of the network. But interestingly, the conversion from PyTorch to ONNX already inserts some errors: this may be caused by differences in the implementation of the operators between PyTorch and ONNX.

\subsection{HW tuning}
\label{section:hw_tuning}

To be able to offload the execution of a CNN on the Gemmini accelerator, C code needs to be generated for each layer of the network. \cite{peccia2022integration} provided an initial approach to integrate the Gemmini accelerator with TVM's microTVM C code generation workflow\footnote{\url{https://github.com/apache/tvm/pull/13770}}. This work presented how the TVM compiler can be used to generate calls to the Gemmini CISC-type and RISC-type instructions, and how it can be used to explore different scheduling candidates.

In this work, we expanded the integration presented in \cite{peccia2022integration} to support the deployment of convolutions, max pooling, resize and concatenation layers using RISC-type instructions on the Gemmini accelerator. This allowed us to use the AutoTVM \cite{autotvm} feature of TVM to optimize them (see Section \ref{section:eval_tuning}). 

\subsection{Model partitioning}
\label{section:model_partitioning}

\begin{table}[t]
	\caption{Gemmini configuration parameters}
	\label{tab:gemmini_params}
	\begin{center}
		\begin{tabular}{ccc} \toprule
			Parameter & Default & Ours \\
			\midrule
			PEs & $16\times16$ & $32\times32$ \\
			Dataflow & Both & Weight Stationary \\
			Scratchpad capacity [KiB] & 256 & 512 \\
			Accumulator capacity [KiB] & 64 & 128 \\
			Scratchpad ports & 1 & 2 \\
			Scratchpad read delay & 4 & 8 \\
			Spatial array output bits & 20 & 18 \\
			Max. in flight mem. requests & 16 & 32 \\
			\bottomrule
		\end{tabular}
	\end{center}
\end{table}

As presented in Section \ref{section:frameworks}, after the quantization process the model can be separated into two clearly distinct parts: the main part, quantized to \textit{int8} representation, and the second part (the post-processing of the bounding boxes using the NMS algorithm) using floating point representation. The first one is perfectly suitable to be executed on the Gemmini accelerator, but the second one does not contain operations that can be offloaded to the accelerator. Given that the FPGA design implemented on the PL of the Zynq architecture runs at a lower frequency than the PS part, it makes sense to run this second part of the model on the PS side. This can be achieved using the TVM framework, by analysing the operator graph of the CNN and separating the model into two parts based on the data type used on each of them (see Section \ref{section:eval_part}).

\section{Evaluation}
\label{section:eval}

For evaluation purposes, we compared the performance of the system implemented on a Xilinx ZCU102 and ZCU111 FPGAs against other hardware, from server-size GPUs (NVIDIA GTX1080), embedded GPU (NVIDIA Jetson AGX Xavier), ARM-based CPUs (Raspberry Pi 4, Quadcore located on the PS of the UltraScale architecture) and another FPGA accelerator (VTA \cite{vta}, implemented on a Xilinx ZCU111 FPGA). All measurements were carried out by tuning and compiling the selected CNN using the TVM framework. 

We also used the original, unmodified Gemmini implemented on a ZCU102 as a baseline. Table \ref{tab:gemmini_params} shows the default parameters of the Gemmini accelerator and the ones we used for our work (we only show the ones that were modified). Table \ref{tab:hw_comparison} shows the resource consumption of all implemented FPGA accelerators. One important aspect to highlight is that, although our optimized Gemmini uses 4 times more PEs than the original one, the amount of used DSPs in the entire design is not even doubled. This demonstrates the effectivity of the DSP packing technique described in Section \ref{section:FPGA_opt}.

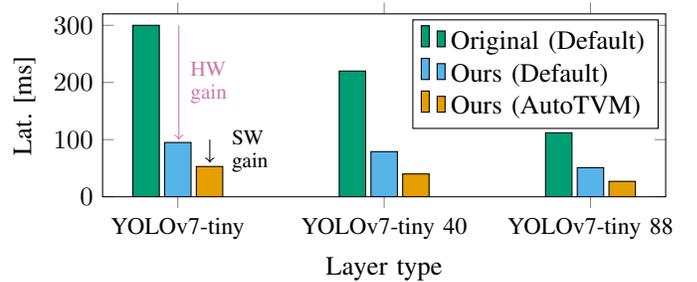
\begin{figure}[t]
	\centering
	\begin{tikzpicture}
		\begin{axis}[
			xlabel=Layer type,
			ylabel={Lat. [ms]},
			enlarge x limits={abs=1cm},
			height=0.17\textheight,
			width=0.5\textwidth,
			ybar,
			ymin=0,
			ymax=320,
			xtick=data,
			cycle list name=ColorBlindFriendlyCycleListBar,
			legend style={at={(0.55,0.97)},anchor=north west, legend columns = 1},
			legend cell align={left},
			symbolic x coords={
				Yolov7-tiny,
				Yolov7-tiny 40,
				Yolov7-tiny 88
			},
			xtick={Yolov7-tiny,Yolov7-tiny 40,Yolov7-tiny 88},
			every x tick label/.append style={font=\small},
			xticklabels={YOLOv7-tiny,YOLOv7-tiny 40,YOLOv7-tiny 88}
			]
			\addplot+ plot table[x=Model version,y=Original,col sep=comma] {./data/gemmini_autotvm.csv};
			\addplot+ plot table[x=Model version,y=Default,col sep=comma] {./data/gemmini_autotvm.csv};
			\addplot+ plot table[x=Model version,y=AutoTVM,col sep=comma] {./data/gemmini_autotvm.csv};
			\addlegendentry{Original (Default)}
			\addlegendentry{Ours (Default)}
			\addlegendentry{Ours (AutoTVM)}
			\node[font=\footnotesize, color=my_purple] at (15,200) {\begin{tabular}{l} HW \\ gain \end{tabular}};
			\node[font=\footnotesize] at (35,80) {\begin{tabular}{l} SW \\ gain \end{tabular}};
			\draw[->, color=my_purple](0.5, 300)--(0.5, 100);
			\draw[->](15.5, 100)--(15.5, 60);
		\end{axis}
	\end{tikzpicture}
	\caption{AutoTVM convolution total latency per model version for the Gemmini accelerator (other kind of layers present a similar behaviour).}
	\label{fig:autotvm_per_model}
\end{figure}

\subsection{Autotuning for the Gemmini accelerator}
\label{section:eval_tuning}

\begin{table*}[t]
	\caption{Energy efficiency of the evaluated hardware platforms (energy is reported per inference)}
	\begin{center}
		\begin{tabular}{ccccccc} \toprule
			HW & \multicolumn{2}{c}{YOLOv7-tiny} & \multicolumn{2}{c}{YOLOv7-tiny 40} & \multicolumn{2}{c}{YOLOv7-tiny 88} \\
			& Energy [J] & Efficiency [GOP/s/J] & Energy [J] & Efficiency [GOP/s/J] & Energy [J] & Efficiency [GOP/s/J] \\
			\midrule
			NVIDIA GTX1080 & 4.58 & 1.68 & 3.28 & 1.49 & 1.78 & 0.95 \\
			NVIDIA Jetson AGX Xavier & 1.89 & 4.06 & 1.31 & 3.72 & 0.72 & 2.36 \\
			ZCU102-Gemmini (Original) & 0.98 & 7.89 & 0.76 & 6.44 & 0.43 & 3.99\\
			\textbf{ZCU102-Gemmini (Ours)} & \textbf{0.28} & \textbf{27.8} & \textbf{0.22} & \textbf{22.12} & \textbf{0.16} & \textbf{10.4} \\
			ZCU111-Gemmini & 0.36 & 21.4 & 0.28 & 17.2 & 0.21 & 7.9 \\
			ZCU111-VTA & 1.89 & 4.07 & 1.57 & 3.13 & 1.03 & 1.65 \\
			\bottomrule
		\end{tabular}
		\label{tab:hw_efficiency}
	\end{center}
\end{table*}

Figure \ref{fig:autotvm_per_model} presents the improvements achieved by executing AutoTVM for convolutions on the Gemmini accelerator. "Default" represents the total latency using the CISC-type instructions provided by the Gemmini developers (as explained in Section \ref{section:hw_tuning}), while "AutoTVM" represents the latency using the best schedules found by the AutoTVM process. When the schedule using RISC-type instructions is not as good as the default one, we default to the CISC-type schedules, to always use the best schedule available.

First of all, our solution achieves a mean speedup of 60 \% when compared with the original unmodified Gemmini (both using the default schedules provided by the Gemmini developers). This demonstrates the importance of our modifications and adaptations for FPGA mapping of the Gemmini accelerator, which allowed us to increase the frequency and the number of processing elements. When analysing the impact of the autotuning process, we are able to achieve a mean 50 \% improvement across all models in the latency of the convolutions. For each model, more than 60 \% of the convolution layers were improved after tuning.  

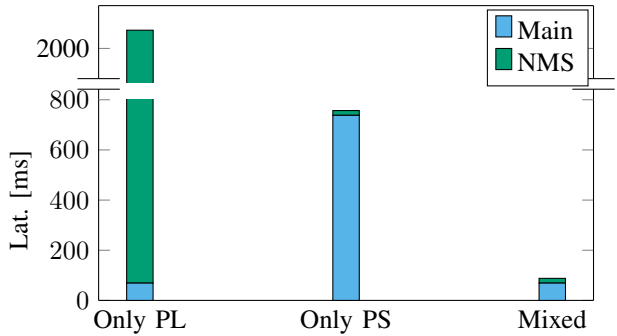
\begin{figure}[t]
	\centering
	\begin{tikzpicture}
		\begin{groupplot}[
			ybar stacked,
			xtick=data,
			width=0.45\textwidth,
			height=0.3\textheight,
			group style={
				group size=1 by 2,
				xticklabels at=edge bottom,
				vertical sep=0pt,
			},
			symbolic x coords={
				Only PL,
				Only PS,
				Mixed
			},
			xtick={
				Only PL,
				Only PS,
				Mixed
			},
			xticklabels={
				Only PL,
				Only PS,
				Mixed
			}
			]
			
			\nextgroupplot[
			height=0.12\textheight,
			scaled y ticks=false,
			axis x line*=top,
			xtick={\empty},
			ytick={2000},
			yticklabels={2000},
			axis y discontinuity=parallel,
			bar shift=-2.74cm,
			ymin=1400,ymax=2500,
			]
			
			\addplot[fill=my_green] plot table[x=option,y=latency,col sep=comma] {./data/di_nms.csv};
			\clip
			(axis cs: {[normalized]-1}, 1600)
			rectangle
			(axis cs: {[normalized]2}, 2500);
			
			\nextgroupplot[
			height=0.18\textheight,
			ymin=0,ymax=800,
			axis x line*=bottom,
			xtick pos=bottom,
			ylabel={Lat. [ms]},
			symbolic x coords={
				Only PL,
				Only PS,
				Mixed
			},
			xtick={
				Only PL,
				Only PS,
				Mixed
			},
			xticklabels={
				Only PL,
				Only PS,
				Mixed
			},
			legend style={at={(0.99,1.45)}, legend columns=1,
				legend cell align={left},}
			]
			\addplot[fill=my_blue] plot table[x=option,y=latency,col sep=comma] {./data/di_main.csv};
			\addplot[fill=my_green] plot table[x=option,y=latency,col sep=comma] {./data/di_nms.csv};
			\addlegendentry{Main}
			\addlegendentry{NMS}
			
		\end{groupplot}
	\end{tikzpicture}
	\caption{Execution of each part of the model on our proposed Zynq-based architecture for the non-pruned YOLOv7-tiny model (pruned models expose the same behaviour).}
	\label{fig:model_partitioning}
\end{figure}

\subsection{Model partitioning}
\label{section:eval_part}

To verify the claim presented in Section \ref{section:model_partitioning}, we executed each part of the model on both PS and PL (Figure \ref{fig:model_partitioning}, reporting already \textit{autotuned} results). The faster execution time for the main part of the model is achieved when tuned for execution on the Gemmini accelerator (PL). The faster execution time for the post-processing is achieved when executing this part of the model on the ARM CPUs (PS). This part takes a lot of time to run on the PL because it does not contain operators that can be offloaded to the accelerator, and the PL is running at a much lower frequency than the PS. Finally, the best solution is to execute the main part on the PL side and the post-processing on the PS side (mixed deployment scenario). Of course, the cost of moving data from the PL to the PS also needs to be taken into account when executing the model in such a distributed way, but because we use shared memory to communicate this data through the ACP port of the Zynq architecture, the cost is negligible and can be ignored.

\subsection{Comparison with other hardware platforms}

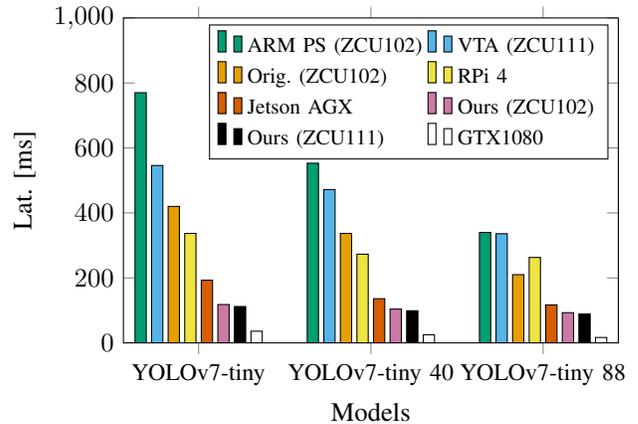
\begin{figure}[t]
	\centering
	\begin{tikzpicture}
		\begin{axis}[
			xlabel=Models,
			ylabel={Lat. [ms]},
			enlarge x limits={abs=1cm},
			height=0.25\textheight,
			width=0.45\textwidth,
			ybar,
			bar width=0.15cm,
			ymin=0,
			ymax=1000,
			xtick=data,
			cycle list name=ColorBlindFriendlyCycleListBar,
			legend style={font=\footnotesize, legend columns=2},
			legend cell align={left},
			symbolic x coords={
				Yolov7-tiny,
				Yolov7-tiny 40,
				Yolov7-tiny 88
			},
			xtick={
				Yolov7-tiny,
				Yolov7-tiny 40,
				Yolov7-tiny 88
			},
			every x tick label/.append style={font=\small},
			xticklabels={
				YOLOv7-tiny,
				YOLOv7-tiny 40,
				YOLOv7-tiny 88
			}
			]
			\addplot+ plot table[x=Model,y=ARM PS (ZCU102),col sep=comma] {./data/hw_autotvm_merged_v2.csv};
			\addplot+ plot table[x=Model,y=VTA (ZCU111),col sep=comma] {./data/hw_autotvm_merged_v2.csv};
			\addplot+ plot table[x=Model,y=Original (ZCU102),col sep=comma] {./data/hw_autotvm_merged_v2.csv};
			\addplot+ plot table[x=Model,y=RPi 4,col sep=comma] {./data/hw_autotvm_merged_v2.csv};
			\addplot+ plot table[x=Model,y=Jetson AGX,col sep=comma] {./data/hw_autotvm_merged_v2.csv};
			\addplot+ plot table[x=Model,y=Ours (ZCU102),col sep=comma] {./data/hw_autotvm_merged_v2.csv};
			\addplot+ plot table[x=Model,y=Ours (ZCU111),col sep=comma] {./data/hw_autotvm_merged_v2.csv};
			\addplot+ plot table[x=Model,y=GTX1080,col sep=comma] {./data/hw_autotvm_merged_v2.csv};
			
			\addlegendentry{ARM PS (ZCU102)}
			\addlegendentry{VTA (ZCU111)}
			\addlegendentry{Orig. (ZCU102)}
			\addlegendentry{RPi 4}
			\addlegendentry{Jetson AGX}
			\addlegendentry{Ours (ZCU102)}
			\addlegendentry{Ours (ZCU111)}
			\addlegendentry{GTX1080}
		\end{axis}
	\end{tikzpicture}
	\caption{Comparison of our proposed platform against other hardware, measured using the TVM framework}
	\label{fig:hw}
\end{figure}

TVM allows us to target different hardware with the exact same model. First, Figure \ref{fig:hw} compares the latency of our proposed platform against other embedded hardware, including a server-size GPU as reference. All measurements were tuned using AutoTVM for the particular hardware. Our solutions using the Gemmini accelerator surpass all other embedded hardware in terms of latency.

Then, we compare the energy consumption of each hardware platform which integrates a power measurement device. Table \ref{tab:hw_efficiency} reports real energy per inference measurements and the efficiency (giga operations per seconds per Joule). For each model, our FPGA solutions are more energy efficient than the other embedded solutions.

\begin{figure}[t]
	\begin{tikzpicture}
		\begin{loglogaxis}[
			legend style={at={(0.985,0.35)}},
			legend cell align={left},
			xlabel={Power [W]},
			ylabel={Throughput [GOP/s]},
			ymin=1,
			ymax=10000,
			xmin=1,
			xmax=110,
			]
			\addplot[nodes near coords,
			every node near coord/.append style={anchor=east, font=\footnotesize},
			only marks,
			point meta=explicit symbolic, mark=square*,mark options={my_green}]
			table[meta=label] {
				x y label
				2.9 26 {}
				4 145 {}
				4.46 14 {VTA}
			};
			\addplot[nodes near coords,
			every node near coord/.append style={anchor=south, font=\footnotesize},
			only marks,
			point meta=explicit symbolic, mark=diamond*,mark options={my_blue}]
			table[meta=label] {
				x y label
				27.2 740.0 \cite{liNovelFPGAAccelerator2020}
				5.05 370.6 \cite{limEnergyEfficientYOLOAccelerator2022}
				26.0 304.5 \cite{wangSparseYOLOHardwareSoftware2020}
			};
			\addplot[nodes near coords, 
			every node near coord/.append style={anchor=east, font=\footnotesize},
			only marks,
			point meta=explicit symbolic, mark=*,mark options={my_red}]
			table[meta=label] {
				x y label
				3.68 110 {}
				2.8 18.9 {}
				3.5 84.3 \cite{guoAngelEyeCompleteDesign2018}
				27.8 1662.0 \cite{liuLeveragingFinegrainedStructured2021}
				8.24 460.8 \cite{shiSparseWinogradConvolutional2018}
				9.82 1249.7 \cite{wuNovelLowCommunicationEnergyEfficient2018}
			};
			\addplot[nodes near coords,
			every node near coord/.append style={anchor=north west, font=\footnotesize},
			only marks,
			point meta=explicit symbolic, mark=triangle*,mark options={my_purple}]
			table[meta=label] {
				x y label
				10.2 1150 \cite{deng3DVNPUFlexibleAccelerator2021}
				9.621 44.8 \cite{liangInSDLAInSSDDeep2019}
				32.0 1578.0 \cite{liuEfficientAcceleratorDNNBased2019}
				1.896 2.56 \cite{wangPYNQBasedFrameworkRapid2018}
				7.31 124.9 \cite{xiaoZacAutomaticOptimization2019}
				11.35 181.8 \cite{yanFPGAbasedMobileNetAccelerator2021}
				6.7 352.0 \cite{yuLightOPUFPGAbasedOverlay2020}
			};
			\legend{YOLOv7, YOLOv2, VGG, Others}
			\addplot[mark=none] coordinates {(1,1) (1000,1000)};
			\addplot[mark=none] coordinates {(1,10) (1000,10000)};
			\addplot[mark=none] coordinates {(1,100) (1000,100000)};
			\addplot[mark=none] coordinates {(1,1000) (1000,1000000)};
			
			\draw[thick,black]
			(axis cs: 3.8,128) circle[radius=2.5mm];
			\node[font=\footnotesize] at (axis cs: 3.8,230) {Ours};
			
			\draw[thick,black]
			(axis cs: 2.8,23) circle[radius=2.5mm];
			\node[font=\footnotesize] at (axis cs: 2.8,42) {Original};
			
			\addplot[dashed, my_red] coordinates {(0.4,14.5) (4000,145000)};
			
			\node[rotate=23, font=\footnotesize] at (axis cs: 65,80) {1 GOP/s/W};
			\node[rotate=23, font=\footnotesize] at (axis cs: 65,800) {10 GOP/s/W};
			\node[rotate=23, font=\footnotesize] at (axis cs: 2,250) {100 GOP/s/W};
			\node[rotate=23, font=\footnotesize] at (axis cs: 2,2500) {1000 GOP/s/W};
			\node[rotate=23, font=\scriptsize,my_red] at (axis cs: 65,3000) {36.5 GOP/s/W};
			
		\end{loglogaxis}
	\end{tikzpicture}
	\caption{Power efficiency of different \textit{int8} CNN accelerators implemented on FPGA (left-up is better).}
	\label{fig:fpga_works}
\end{figure}
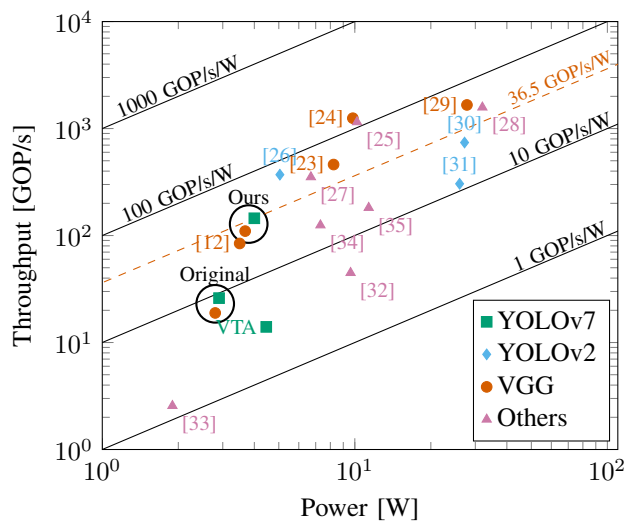

Finally, Figure \ref{fig:fpga_works} compares our solutions against other FPGA works that implemented accelerators for \textit{int8} quantized CNNs. First, our solution improves the efficiency of the Gemmini accelerator when compared against the unmodified one. Second, our work is the first one of the compared works to execute a YOLOv7 model version on an FPGA, which because of the amount of layers of the model (58 convolution layers) prevents it from being implemented using a stream-type accelerator. Third, our solution lies on the Pareto-optimal border of designs. Finally, it can be appreciated that our Gemmini-based solution achieves comparable or better power efficiency than other works. Those achieving an efficiency greater than 36.5 GOP/s/W can be explained because they designed specialized accelerators to execute the Winograd convolution \cite{shiSparseWinogradConvolutional2018,wuNovelLowCommunicationEnergyEfficient2018,deng3DVNPUFlexibleAccelerator2021}, or used higher working frequencies (200 MHz \cite{limEnergyEfficientYOLOAccelerator2022,yuLightOPUFPGAbasedOverlay2020, liuEfficientAcceleratorDNNBased2019}, 242 MHz \cite{liuLeveragingFinegrainedStructured2021}). Both could be explored in a future work to improve the efficiency of our solution.






\section{Case study}
\label{section:case_study}

\begin{figure}[htbp]
	\centering
	\includegraphics[width=0.35\textwidth, angle=-90]{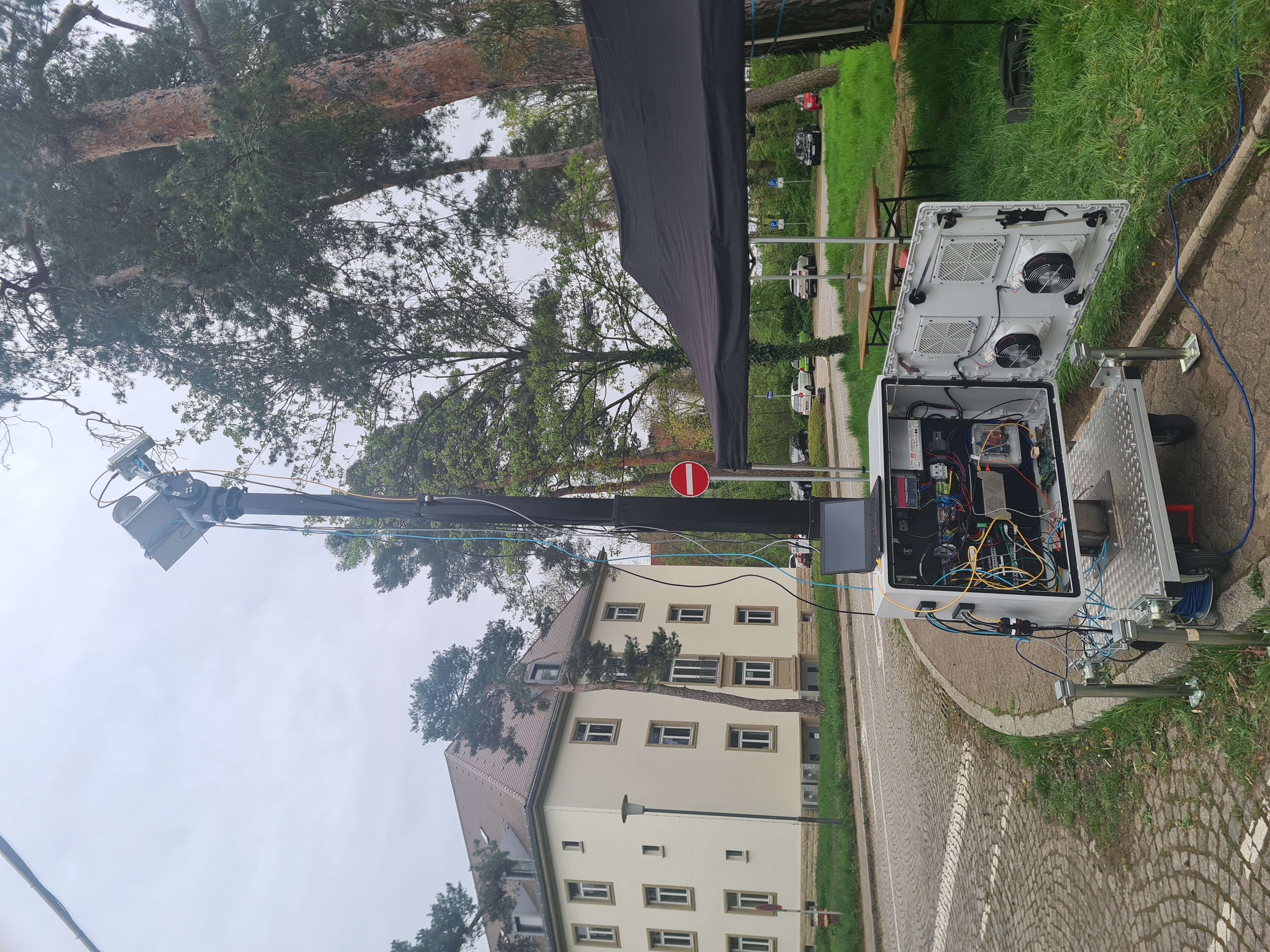}
	\caption{The Infra2Go platform in the intersection monitoring scenario of our case study.}
	\label{fig:packaging}
\end{figure}

A CNN model cannot exist in isolation: the integration of standard interfaces to connect to cameras and publish the results presents its own challenges. This is why few FPGA CNN papers demonstrate how to integrate their solution into a larger system. As a case study, traffic surveillance presents itself as an interesting use case for implementation of realtime edge systems \cite{10293592}. We used the mobile sensor platform Infra2Go \cite{fleck2022infra2go} to integrate our FPGA-based solution and realize this use case (Figure \ref{fig:packaging}).

The proposed FPGA accelerator is integrated into a larger processing pipeline using ROS2 \cite{ros2} as a middleware: \begin{inparaenum}[1)]
	\item camera images of a calibrated camera are send from an x86 based host ECU to the FPGA in form of standardized ROS2 messages over ethernet,
	\item the Zephyr real-time operating system \cite{zephyr} is used as the runtime to receive the image, execute the main part of the model on the RISC-V core and the Gemmini accelerator,
	\item the TVM runtime on the PS receives the intermediate outputs, runs the NMS post-processing and publishes the object detections, and
	\item on the main ECU, homography projection to a ground plane, world-space tracking and state estimation (including velocity estimation) are performed using a Gaussian Mixture Probability Hypothesis Density Filter (GMPHD).
\end{inparaenum}

\section{Conclusion}
\label{section:conclusions}

This paper presented a deployment framework for quantized CNNs on an FPGA system build around the Gemmini accelerator. We modified the Gemmini design to enable improved mapping onto an FPGA, which resulted in a mean speed-up of 60 \% when compared against the original, unmodified Gemmini. Additionally, the autotuning process allowed us a further 50 \% improvement across all model versions. The evaluation also demonstrated the effectiveness of model partitioning for the overall model execution including post-processing of the detected bounding boxes, a process usually neglected when deploying CNNs on FPGAs. Real measurements and comparative analysis against other hardware platforms revealed that the proposed FPGA solution outperforms all others in terms of energy efficiency, achieving a 85\% reduction when compared with an NVIDIA Jetson AGX Xavier board, and 93\% when compared against a server-size GPU. Furthermore, our Gemmini-based solution achieves a power efficiency of 36.5 GOP/s/W, which is better or comparable to the same metric reported by other similar FPGA accelerator works. Finally, we showcased using a traffic monitoring use case how this kind of solution can be integrated into a wider system.

	\balance
	
	\bibliographystyle{IEEEtran}
	\bibliography{bib}
	
	
\end{document}